# Strain induced incommensurate phases in hexagonal manganites

Fei Xue,[1,*] Xueyun Wang,[2,3] Sang-Wook Cheong,[3] and Long-Qing Chen[1]

[1]Department of Materials Science and Engineering, The Pennsylvania State University, University Park, Pennsylvania 16802, USA

[2]School of Aerospace Engineering, Beijing Institute of Technology, Beijing 100081, China

[3]Rutgers Center for Emergent Materials and Department of Physics and Astronomy, Rutgers University, Piscataway, NJ 08854, USA

*Corresponding author: xuefei5376@gmail.com

**Abstract**:

An incommensurate phase refers to a solid state in which the period of a superstructure is incommensurable with the primitive unit cell. Recently the incommensurate phase is induced by applying an in-plane strain to hexagonal manganites, which demonstrates single chiral modulation of six domain variants. Here we employ Landau theory in combination with the phase-field method to investigate the incommensurate phase in hexagonal manganites. It is shown that the equilibrium wave length of the incommensurate phase is determined by temperature and the magnitude of the applied strain, and a temperature-strain phase diagram is constructed for the stability of the incommensurate phase. Temporal evolution of domain structures reveals that the applied strain not only produces the force pulling the vortices and anti-vortices in opposite directions, but also results in the creation and annihilation of vortex-antivortex pairs.

# I. Introduction

The phase transitions in solids include two types of significant transitions, i.e., order-disorder and displacive transitions [1]. The extra order and relative atomic displacement may give rise to a superstructure, with additional reflection spots in the diffraction pattern compared to that from a primitive unit cell. If the indices of the additional reflection spots are all rational numbers, the phase is called a commensurate (C) phase [2]. For example, the reflection spot with index (1/2, 0, 0) indicates that the there exists a superstructure caused by unit cell doubling along the first axis. On the other hand, if at least one index of the additional reflection spots is an irrational number, the system totally loses its translational symmetry along a particular direction, and the phase is called an incommensurate (IC) phase [3]. The IC phase is found to play a significant role in IC dielectrics, IC magnetics, charge-density wave systems, and spin-Pierls compounds [4-7]. In ferroelectrics, the IC phase can be induced by size effect [8], flexoelectric effect [9], and applied strain [10].

Hexagonal ($h$-) REMnO$_3$ (RE, rare earths) have recently attracted enormous attention due to their intrinsic multiferroic properties and intriguing domain patterns [11,12]. REMnO$_3$ are improper ferroelectrics in which the polarization is induced by the structural trimerization during the transition from space group *P6$_3$/mmc* to space group *P6$_3$cm* [13-15]. The structural trimerization results in three translational phase variants based on the different choices of the origin, and each variant has two options of polarization, i.e., either along +$c$ or –$c$ directions [11,16]. Thus, there are totally six C domain variants in the REMnO$_3$ systems. The six types of domains can cycle around the vortex and anti-vortex cores resulting in vortex domains, as shown in Figs. 1(a) and 1(b) [11,12]. Another type of domain structures in REMnO$_3$ is the single-chirality striped domains with fixed sequence of the six domains (Figs. 1c and 1d), which is caused by the coupling between an in-plane strain and the gradient of the trimerization [10,16]. The modulation of the six domain variants leads to a superstructure and the length ratio between the superstructure and the primitive unit cell can be an irrational number, i.e., the single-chirality striped domains can be treated as an IC phase.

In this paper, we employ Landau theory and the phase-field method to investigate the stability and properties of the IC phase in $h$-REMnO$_3$. Compared with the analytical results of the classical XY model, the effect of the energy anisotropy in the order parameter space of $h$-REMnO$_3$ is demonstrated. Surprisingly, an unusual local enhancement of the trimerization is observed when the applied strain is large. The equilibrium wave length of the IC phase is found to be determined by temperature and the magnitude of the applied strain, and a temperature-strain phase diagram is constructed for the stability of the IC phase. It is shown that the applied in-plane strain gives rise to an IC phase at high temperatures, which is frozen by limited domain wall mobility at low temperatures. Phase-field simulations demonstrate that the evolution from vortex domains to the IC phase is caused not only by the antiparallel movement of vortex and anti-vortex cores, but also by the creation and annihilation of vortex-antivortex pairs.

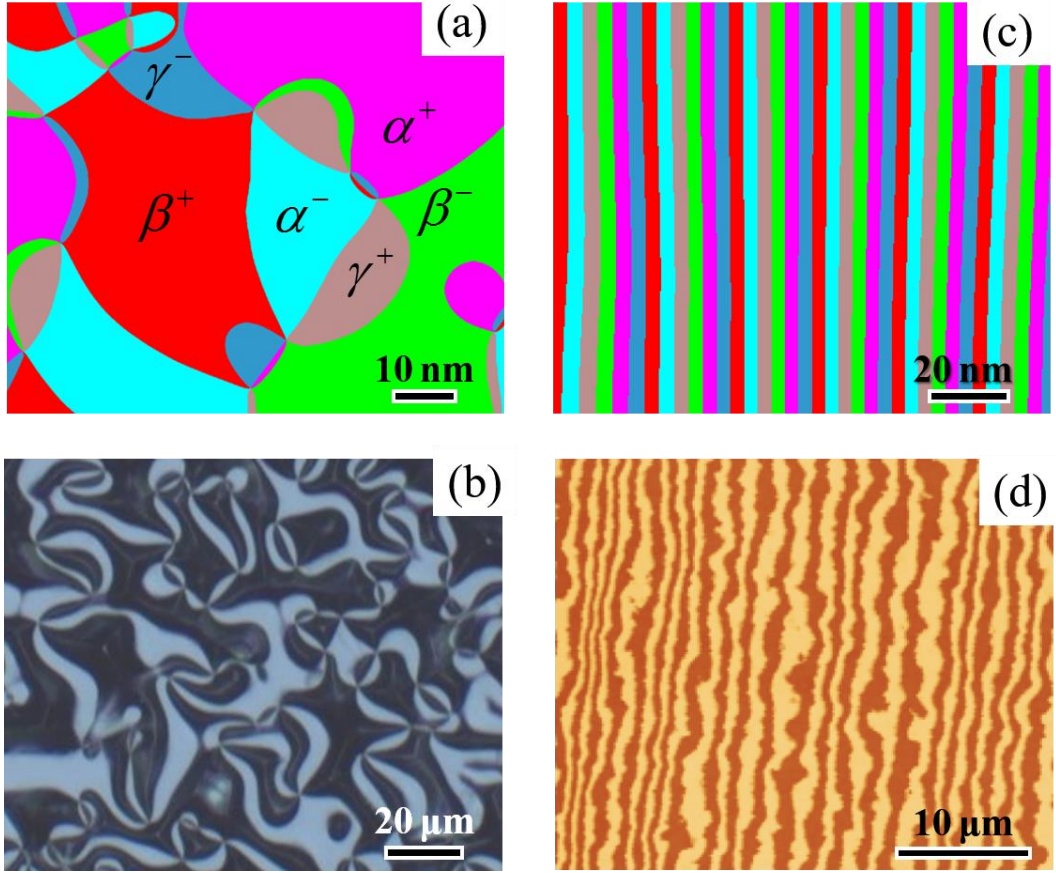

FIG. 1. Domain patterns on the basal plane of $h$-REMnO$_3$. (a) and (b) Vortex domains from (a) phase-field simulation and (b) optical microscope. (c) and (d) Single-chirality striped domains from (c) phase-field simulation and (d) atomic force microscope scanning on a chemically etched sample.

## II. Results and discussion

### A. Incommensurate phase in the XY model

We start our discussion from the IC phases in the classical XY model. In the XY model, the order parameter is a two-dimensional (2D) vector, which can be described by the magnitude $Q$ and phase $\phi$ in the polar coordinate system. The energy density is a function of $Q$ and independent of $\phi$, i.e., the energy density is isotropic in the order parameter space (Fig. 2a) [17]. Assuming that there exists an order parameter modulation along the $x$ direction, the free energy density describing the IC phases is given by [2],

$$f_{XY} = \frac{a}{2}Q^2 + \frac{b}{4}Q^4 + \frac{1}{2}g(\frac{\partial Q}{\partial x})^2 + \frac{1}{2}gQ^2(\frac{\partial \phi}{\partial x})^2 + \xi Q^2 \frac{\partial \phi}{\partial x}, \quad (1)$$

where $a$ and $b$ are the coefficients of Landau polynomial, $g$ is the coefficient of gradient energy, the term $\xi Q^2 \frac{\partial \phi}{\partial x}$ is called the Lifshitz invariant [2], and $\xi$ is the corresponding coefficient.

The evolution of the system is described by the Euler–Lagrange equations,

$$\frac{\delta f_{XY}}{\delta Q} = bQ^3 + Q[a + 2\xi\frac{\partial \phi}{\partial x} + g(\frac{\partial \phi}{\partial x})^2] - g\frac{\partial^2 Q}{\partial x^2} = 0, \quad (2)$$

$$\frac{\delta f_{XY}}{\delta \phi} = -Q[2\frac{\partial Q}{\partial x}(\xi + g\frac{\partial \phi}{\partial x}) + gQ\frac{\partial^2 \phi}{\partial x^2}] = 0, \quad (3)$$

The solution of Eqs. (2) and (3) is

$$Q = \sqrt{-\frac{a}{b} + \frac{\xi^2}{bg}}, \quad \phi = C_1 x + C_2, \quad (4)$$

where $C_1$ and $C_2$ are the constants of integration. Without lost of generality, let $\phi = 0$ at $x = 0$, and we obtain $C_2 = 0$, i.e.,

$$Q = \sqrt{-\frac{a}{b} + \frac{\xi^2}{bg}}, \quad \phi = C_1 x, \quad (5)$$

Equation (5) indicates that the value of $Q$ is larger than or equal to that with $\xi = 0$, since $Q \geq Q_0 = \sqrt{-\frac{a}{b}}$. Thus a nonzero Lifshitz invariant term increases the magnitude of $Q$. In Eq. (1), only $a$ is dependent on temperature $T$, i.e., $a = a_0(T - T_C)$, where $a_0$ is a constant and $T_C$ is the Curie temperature. By solving $Q=0$ in Eq. (5), we obtain the transition temperature

$$T_0 = T_C + \frac{\xi^2}{a_0 g}, \tag{6}$$

Therefore, Eq. (1) with $\xi = 0$ describes a second-order phase transition with transition temperature $T_0 = T_C$, and a nonzero Lifshitz invariant term increases the transition temperature.

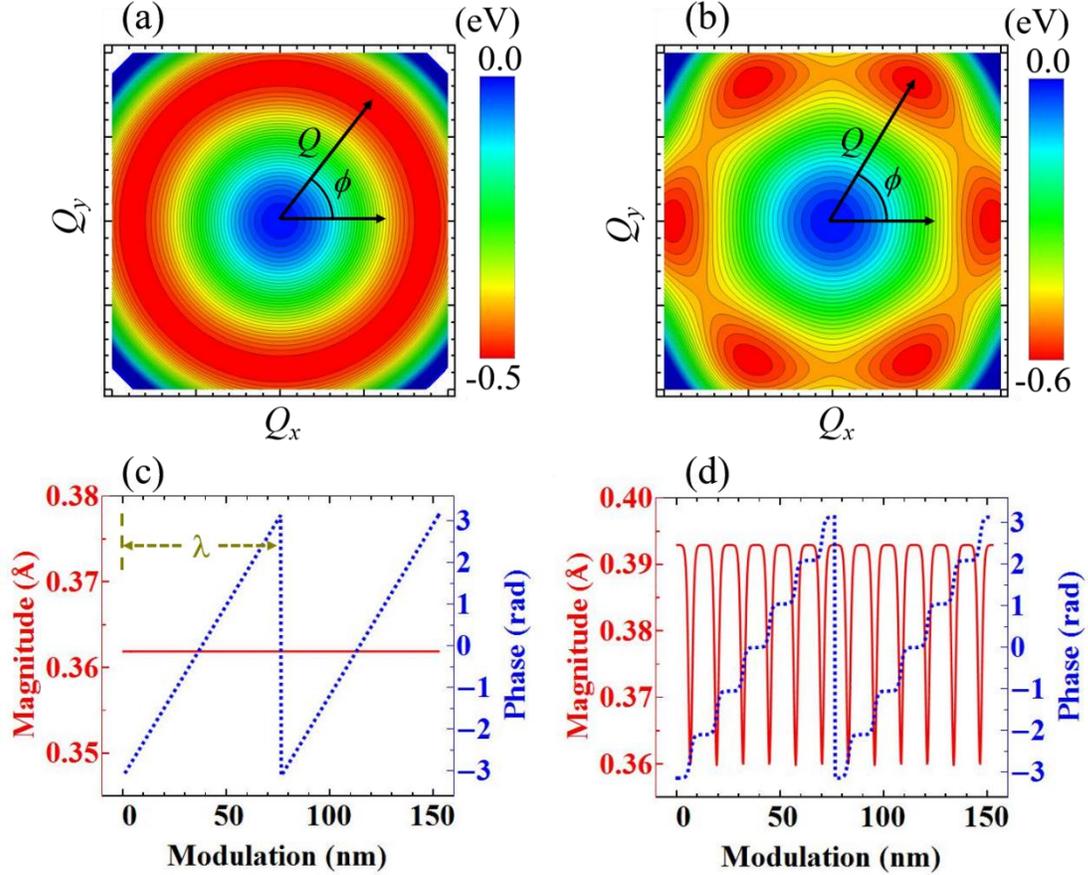

FIG. 2. Comparison between the XY model and YMO. (a) and (b) Energy landscape in the order parameter space for (a) the XY model and (b) YMO. The magnitude and phase of the order parameter are labelled by $Q$ and $\phi$, respectively. (c) and (d) Distribution of the order parameter

along the modulation direction for (c) the XY model and (d) YMO. In (c), λ denotes the wave length of the modulation.

Next phase-field simulations based on a semi-implicit spectral method are employed to numerically solve the solution of Eq. (1) [18,19]. In the phase-field simulations, the polar coordinates $Q$ and $\phi$ are transformed into Cartesian coordinates $(Q_x, Q_y)$, with $Q_x = Q\cos\Phi, Q_y = Q\sin\Phi$ [20]. Then Eq. (1) becomes

$$f = \frac{a}{2}(Q_x^2 + Q_y^2) + \frac{b}{4}(Q_x^2 + Q_y^2)^2 + \frac{g}{2}[(\frac{\partial Q_x}{\partial x})^2 + (\frac{\partial Q_y}{\partial x})^2] + \xi(Q_x \frac{\partial Q_y}{\partial x} - Q_y \frac{\partial Q_x}{\partial x}), \qquad (7)$$

The coefficients $a$, $b$, and $g$ use the same values with those of YMnO$_3$ (YMO), which are obtained from first-principles calculations at 0 K [16], and it is assumed that $a = a_0(T - T_C)$ with $T_C = 1200$ K. The solution of Eq. (7) is obtained by numerically solving the time-dependent Ginzburg-Landau (TDGL) equations

$$\frac{\delta Q_x}{\delta t} = L_Q \frac{\delta f}{\delta Q_x}, \frac{\delta Q_y}{\delta t} = L_Q \frac{\delta f}{\delta Q_y}, \qquad (8)$$

where $L_Q$ is the kinetic coefficient related to the domain wall mobility. The system size is $1024\Delta x \times 1\Delta x \times 1\Delta x$ with $\Delta x = 0.30\ nm$. Periodic boundary conditions are applied to the system along three dimensions.

Figure 2(c) presents the result from a phase-field simulation with $T=1000$ K and $\xi = 0.75\ eV/nm$. The result is consistent with Eq. (6), i.e., the magnitude $Q$ is a constant and the phase $\phi$ can be transformed to be a linear function of the spatial coordinate. In Fig. 2(c), the value of $\phi$ is chosen to be between $-\pi$ to $\pi$, and $\phi$ becomes a periodic function of the spatial coordinate. The period is the wave length of the modulation, labelled as $\lambda$ in Fig. 2(c).

## B. Incommensurate phase in hexagonal REMnO$_3$

Here we employ $h$-YMO as an example of $h$-REMnO$_3$. In YMO, the primary order parameters $Q$ and $\phi$ characterize the structural trimerization, and the energy density possesses six-fold anisotropy in the order parameter space (Fig. 2b), in contrast to the isotropic energy landscape in the XY model (Fig. 2a). Meanwhile, the trimerization induces a secondary order parameter, i.e., polarization $P_z$. The total free energy density of $h$-YMO is given by [16,20,21]

$$
\begin{aligned}
f_{YMO} &= \frac{a}{2}(Q_x^2 + Q_y^2) + \frac{b}{4}(Q_x^2 + Q_y^2)^2 + \frac{c}{6}(Q_x^2 + Q_y^2)^3 + \frac{c'}{6}(Q_x^6 - 15Q_x^4Q_y^2 + 15Q_x^2Q_y^4 - Q_y^6) \\
&- h(Q_x^3 - 3Q_xQ_y^2)P_z + \frac{h'}{2}(Q_x^2 + Q_y^2)P_z^2 + \frac{a_P}{2}P_z^2 + \frac{g}{2}[(\frac{\partial Q_x}{\partial x})^2 + (\frac{\partial Q_x}{\partial y})^2 + (\frac{\partial Q_y}{\partial x})^2 \\
&+ (\frac{\partial Q_y}{\partial y})^2] + \frac{s_P^x}{2}[(\frac{\partial P_z}{\partial x})^2 + (\frac{\partial P_z}{\partial y})^2] - E_zP_z + \frac{s_Q^z}{2}[(\frac{\partial Q_x}{\partial z})^2 + (\frac{\partial Q_y}{\partial z})^2] + \frac{s_P^z}{2}(\frac{\partial P_z}{\partial z})^2 \\
&- \frac{1}{2}\varepsilon_0\kappa_b E_zE_z + G[(\varepsilon_{xx} - \varepsilon_{yy})(Q_x\frac{\partial Q_y}{\partial x} - Q_y\frac{\partial Q_x}{\partial x}) - 2\varepsilon_{xy}(Q_x\frac{\partial Q_y}{\partial y} - Q_y\frac{\partial Q_x}{\partial y})]
\end{aligned}
\quad , \quad (9)
$$

where $c, c', h, h'$, and $a_P$ are the coefficients of Landau polynomial, $s_Q^z, s_P^x$, and $s_P^z$ are the coefficients of gradient energy, $\varepsilon_0$ is the vacuum permittivity, $\kappa_b$ is the background dielectric constant [22], $E_z$ is the electric field calculated by $E_z = -\frac{\partial \varphi}{\partial z}$ with $\varphi$ the electrostatic potential, $\varepsilon_{ij}$ is the strain tensor, and $G$ is the coupling coefficient between the applied strain and the gradient of the trimerization. The coefficients of Landau polynomial and gradient energy are obtained from first-principles calculations at 0 K [16], and it is assumed that $a = a_0(T - T_C)$ with $T_C = 1200$ K. The background dielectric constant $\kappa_b$ takes the typical value of 50 [23]. For simplicity, we assume that $\varepsilon_{xy}=0$ and $\xi = G(\varepsilon_{xx}-\varepsilon_{yy})$. Since $G$ is a constant and $\xi$ is proportional to the applied strain, the energy contribution from the Lifshitz invariant can be controlled by the magnitude of the applied strain.

The system is evolved by numerically solving the TDGL equations

$$\frac{\delta P_z}{\delta t}=L_P\frac{\delta f}{\delta P_z},\ \frac{\delta Q_x}{\delta t}=L_Q\frac{\delta f}{\delta Q_x},\ \frac{\delta Q_y}{\delta t}=L_Q\frac{\delta f}{\delta Q_y}, \tag{10}$$

where $L_P$ is the kinetic coefficient related to the domain wall mobility.

Although the value of $P_z$ is solved in Eq. (10), its specific value is not significant for the IC phase, and thus in this paper we only show the profiles of the primary order parameter, i.e., $Q$ and $\phi$. Figure 2(d) shows the modulation of $Q$ and $\phi$ from a phase-field simulation when $T$=1000 K and $\xi = 0.75\ eV/nm$, the same parameter setting with that of Fig. 2(c). In contrast to the linear distribution of $\phi$ in the XY model, $\phi$ in YMO exhibits staircase-like plateaus at the values of $0,\pm\frac{\pi}{3},\pm\frac{2\pi}{3},\pm\pi$, which correspond to the six energy minima in Fig. 2(b). These plateaus can be treated as six C domains, and the transition regions between neighboring plateaus correspond to domain wall regions in the C phase. Different from being a constant as in Fig. 2(c), $Q$ in Fig. 2(d) takes its minima at the transition region, which is similar to the situation at domain walls of the C phase [16].

### C. Order parameter profiles under different $T$ and $\xi$

In this section we investigate the effect of different $T$ and $\xi$ on the order parameter profiles of YMO. In Figs. 3(a)-3(c), the value of $\xi$ is fixed at $0.75\ eV/nm$, and $T$ is varied. When $T$ equals to 1195 K, just below $T_C$, the distribution of the phase $\phi$ is almost a straight line, close to the situation of the XY model (Fig. 3a). This is because the energy anisotropy is reduced near $T_C$ and the continuous XY model emerges at $T_C$ [24-26]. The magnitude $Q$ changes similar to a sine function with the maxima obtained at phase $\phi = \frac{i\pi}{3} (i = 0-3)$ and minima at phase $\phi = \frac{i\pi}{6} (i = 1, 3, 5)$. Note that the variation of $Q$ at this temperature is small, ~0.3% of the total magnitude. At a lower temperature 1170 K, the phase $\phi$ develops staircase-like plateaus, and near the transition regions, $Q$ shows valley-shape decrease (Fig. 3b). When the temperature is further decreased to 1100 K, the plateaus becomes wider, and the transition regions are narrower (Fig. 3c). Note that the variation of $Q$ at 1100 K is ~6% of the total magnitude, larger than that at 1195 K. Therefore, the profiles of $Q$ and $\phi$ evolve from that of the XY model to that of the C phase with a decreasing temperature.

As illustrated in Figs. 3(a)-3(c), when $\xi$ is small, the profiles of $Q$ are generally as expected, i.e., $Q$ takes its minima at the transition regions. When $\xi$ is large, however, the profiles of $Q$ show unexpected behaviors. When $T$ equals to 1195 K, $Q$ behaviors like a sine function with its minima at phase $\frac{i\pi}{3} (i = 0-3)$ and maxima at phase $\frac{i\pi}{6} (i = 1, 3, 5)$, as plotted in Fig. 3(d). This means that $Q$ at the transition regions is larger than that within a region corresponding to the energy minima in Fig. 2(b). The abnormal enhancement of $Q$ near the transition regions is because when $\frac{dQ}{dx}$ is small and $\xi$ is large, the contribution from the gradient energy $\frac{1}{2} g (\frac{\partial Q}{\partial x})^2$ can be ignored. By neglecting the higher-order terms and the coupling with polarization, Eq. (9) can be approximated as,

$$f_{XY} = [\frac{a}{2} + \xi\frac{\partial\phi}{\partial x} + \frac{1}{2}g(\frac{\partial\phi}{\partial x})^2]Q^2 + \frac{b}{4}Q^4, \tag{11}$$

which leads to $Q = \sqrt{(a + 2\xi\frac{\partial\phi}{\partial x} + g(\frac{\partial\phi}{\partial x})^2)/b}$. Since $\xi$ and $\frac{\partial\phi}{\partial x}$ have the same sign from Eq. (14) as shown below, $Q$ is larger under a larger $\frac{\partial\phi}{\partial x}$, i.e., near the transition regions. When $T$ is decreased to 1170 K, $\frac{dQ}{dx}$ becomes larger, and the contribution from $\frac{1}{2}g(\frac{\partial Q}{\partial x})^2$ cannot be ignored. The profile of $Q$ demonstrates complex behaviors due to the competitions between $\frac{1}{2}g(\frac{\partial Q}{\partial x})^2$ and the Lifshitz invariant $\xi Q^2 \frac{\partial\phi}{\partial x}$. As shown in Fig. 3(e), $Q$ has two types of local minima, i.e., in the middle of a plateau of $\phi$ and in the middle of a transition region. The maxima of $Q$ are obtained near the boundaries of the plateaus. When $T$ is further decreased to 1100 K, $Q$ becomes flat within the plateaus of $\phi$ while small overshoots are observed near the transition regions (Fig. 3f). The overshoots near the transition regions are also reported in earlier references [2,27]. Therefore, when $\xi$ is large, i.e., when the applied strain is large, due to the significant energy contribution from the Lifshitz invariant, the magnitude of the order parameter $Q$ shows abnormal behaviors near the transition regions.

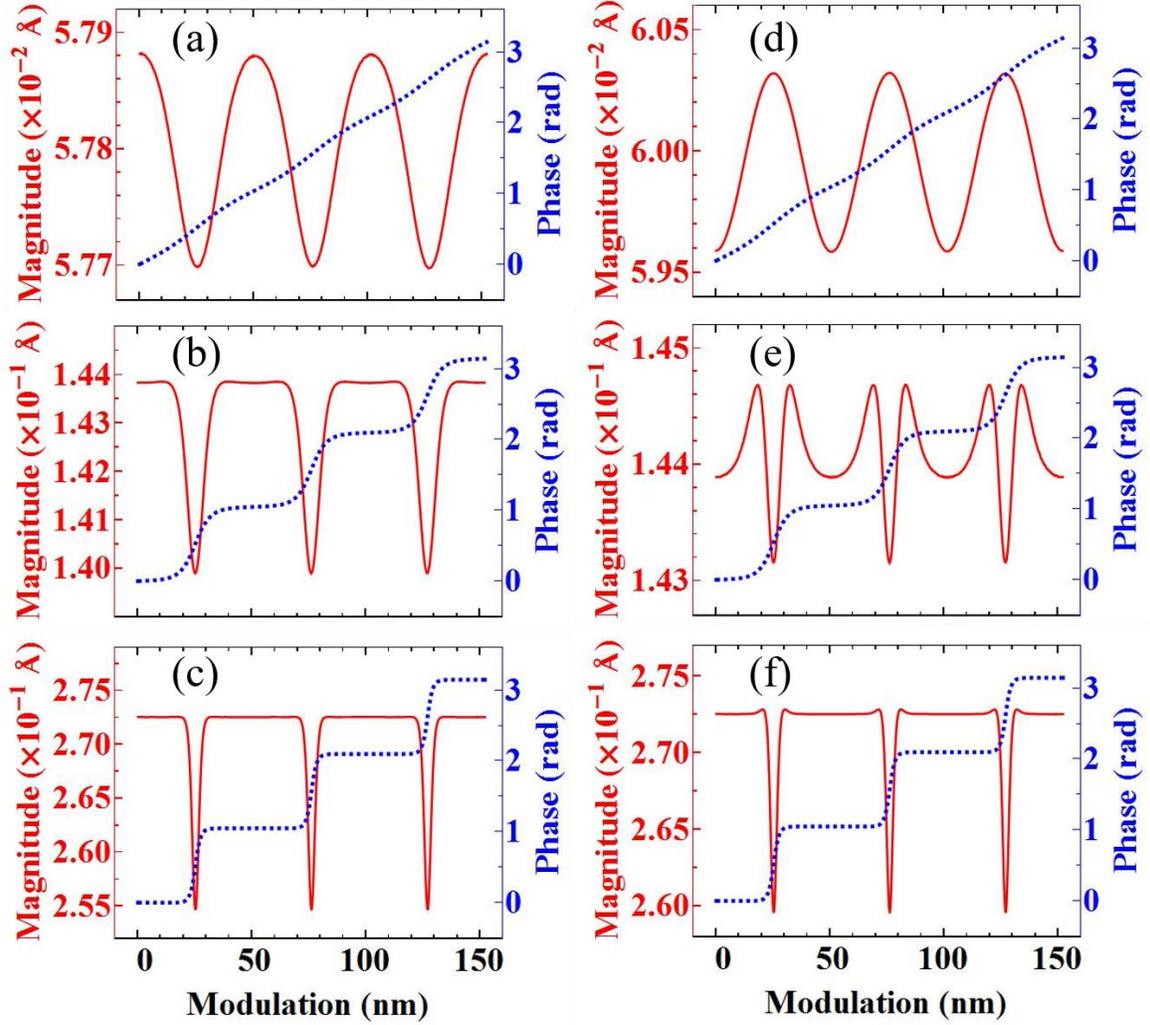

FIG. 3. Profiles of the order parameters $Q$ and $\phi$ in $h$-YMO with different temperatures and strains. (a)-(c) Profiles of $Q$ and $\phi$ with $\xi = 0.75\ eV/nm$ at (a) 1195 K, (b) 1170 K, and (c) 1100 K. (d)-(f) Profiles of $Q$ and $\phi$ with $\xi = 2.83\ eV/nm$ at (d) 1195 K, (e) 1170 K, and (f) 1100 K.

### D. Equilibrium wave length of the incommensurate phase

In this section we investigate the equilibrium wave length of the IC phase, which is labelled as $\lambda_0$. As a limiting case with an isotropic energy landscape, the analytical solution of the XY model is studied first. Substituting Eq. (5) into Eq. (1), we obtain

$$f_{XY} = \frac{\xi^2}{4bg^2} - \frac{a^2}{4b} + \frac{(\xi^2 - ag)}{2bg}(gC_1^2 + 2\xi C_1), \tag{12}$$

From the relation $C_1 = \frac{2\pi}{\lambda}$, the free energy can be written as a function of the wave length,

$$f_{XY} = \frac{\xi^2}{4bg^2} - \frac{a^2}{4b} + \frac{2(\xi^2 - ag)}{bg}[g(\frac{\pi}{\lambda})^2 + \xi\frac{\pi}{\lambda}], \tag{13}$$

$f_{XY}$ as a function of $\lambda$ at 1100 K with $\xi$=2.39 eV/nm is plotted as a pink line in Fig. 4(a), which shows an energy minimum at equilibrium wave length $\lambda_0$. From $\frac{\partial f_{XY}}{\partial \lambda} = 0$ in Eq. (13), $\lambda_0$ is expressed by

$$\lambda_0 = \frac{2\pi g}{\xi}, \tag{14}$$

which is shown as a pink line in Fig. 4(b). As indicated by Eq. (14), $\lambda_0$ in the XY model is a function of $\xi$, and independent of temperature.

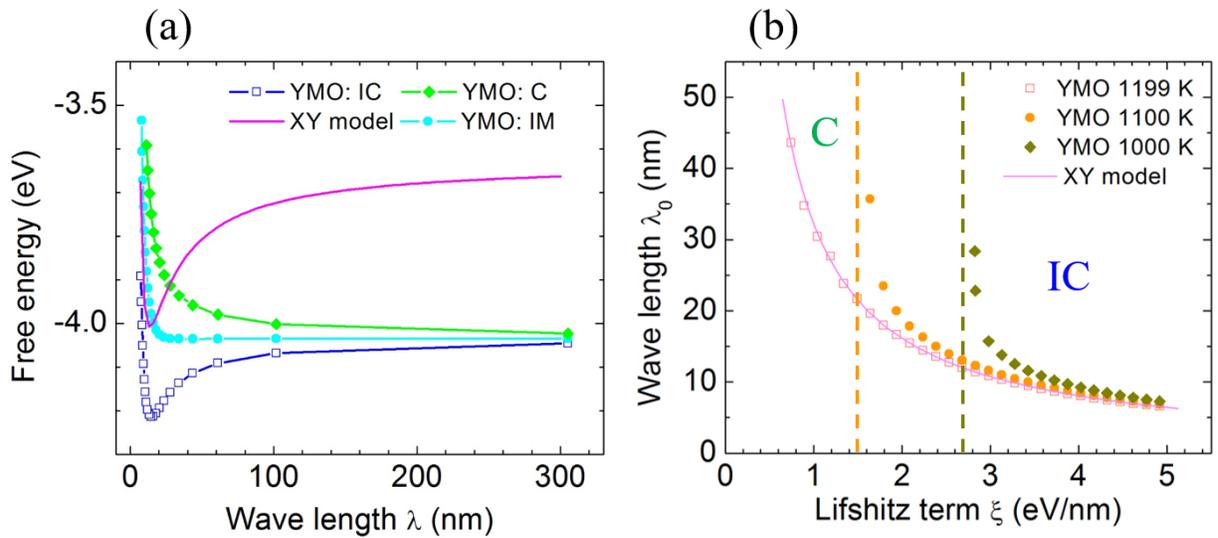

FIG. 4. Equilibrium wave length of the IC phase. (a) Free energy as a function of wave length for the XY model and YMO. The pink line denotes the free energy of the XY model at 1100 K with $\xi$=2.39 eV/nm. The blue line, cyan line, and green line are the results of YMO at 1100 K with $\xi$=2.39 eV/nm, 1.64 eV/nm, and 0.90 eV/nm, respectively, which represent the energy profiles of the IC phase, intermediate state (IM), and C phase, respectively. (b) Equilibrium wave length as a function of $\xi$ for the XY model and YMO at different temperatures.

Then the equilibrium wave length in YMO is investigated. When $\xi$ equals to 2.39 eV/nm at 1100 K, YMO is in the IC phase. As shown in the blue line of Fig. 4(a), the energy profile of YMO is similar to that of the XY model, and there exists an equilibrium wave length $\lambda_0$. Note that in Fig. 3(c), the transition regions between neighboring plateaus can be treated as solitons [2]. The IC phase can be stabilized over the C phase since the self-energy of the solitons is negative [21,28]. On the other hand, the interaction between adjacent solitons is repulsive [2,21]. Therefore, the balance of the two energies gives rise to the equilibrium density of solitons, and consequently equilibrium wave length.

When $\xi$ equals to 0.90 eV/nm at 1100 K, YMO is in the C phase. As shown by the green line in Fig. 4(a), the energy profile is a monotonically decreasing function of the wave length, and the corresponding self-energy of solitons, i.e., the domain wall energy, is positive [21]. The IC and C phases are separated by an intermediate state, and the corresponding energy profile is plotted as the cyan line in Fig. 4(a), where the slop is small for a large $\lambda$. In the intermediate state, the self-energy of solitons is zero, and the energy monotonically decreases due to the repulsive interaction between adjacent solitons [2].

The equilibrium wave length $\lambda_0$ as a function of $\xi$ in YMO at different temperatures is demonstrated in Fig. 4(b). The evolution of $\lambda_0$ at 1199 K is almost the same with that of the XY model, since the energy landscape of YMO near $T_C$ is close to that of the XY model [26]. When $T$ equals to 1100 K, the value of $\lambda_0$ deviates from that of the XY model with a decreasing $\xi$. Eventually when $\xi$ decreases to a critical value, the value of $\lambda_0$ approaches infinity, which corresponds to the intermediate state in Fig. 4(a). With $\xi$ below the critical value, the C phase is stable, and there exists no $\lambda_0$. Also, the equilibrium wave length at 1000 K is plotted as the olive line in Fig. 4(b), which shows a larger critical value of $\xi$ than at 1100 K. Note that $\lambda_0$ is close to that of the XY model when $\xi$ is large for the three temperatures, due to the dominant energy contribution from the Lifshitz invariant.

### E. Temperature-strain phase diagram

By calculating the critical values of $\xi$ at different temperatures, a $T$-$\xi$ phase diagram for the stabilities of the IC and C phases is constructed, as shown in Fig. 5. Based on Eq. (14), when $\xi$ changes its sign, $\lambda_0$ also switches its sign, i.e., the system flips its modulation direction to accommodate the sign change of $\xi$. Therefore, the IC-C boundary is symmetric with respect to $\xi=0$. Also, Fig. 5 shows that the transition temperature from the high-symmetry phase to the IC phase increases with an increasing $\xi$, which is consistent with the conclusion of Eq. (6). Since $\xi = G(\varepsilon_{xx} - \varepsilon_{yy})$, the $T$-$\xi$ phase diagram is also a temperature-strain $T$-$(\varepsilon_{xx} - \varepsilon_{yy})$ phase diagram.

As shown by the IC-C boundary in Fig. 5, the critical value of $\xi$, i.e., the critical strain, increases with a decreasing temperature. We expect that it is challenging to induce the IC phase from the C phase at room temperature by applying strain, since the crystal may break down before the critical strain. Therefore, to induce the C to IC phase transition, we need to anneal the sample to high temperatures and then apply the strain. If the applied strain is fixed, and the temperature is cooling down, the induced IC phase will lose its thermodynamic stability when crossing the IC-C phase boundary in Fig. 5. However, if the domain wall and vortex core mobility is limited at this temperature [24,29], the domain patterns will be frozen, and the striped IC domains can be observed at room temperature. In fact, a single-domain C state possesses the smallest energy in *h*-YMO at room temperature, and both the striped IC domains and vortex domains are frozen by the limited domain wall mobility at low temperatures [29,30].

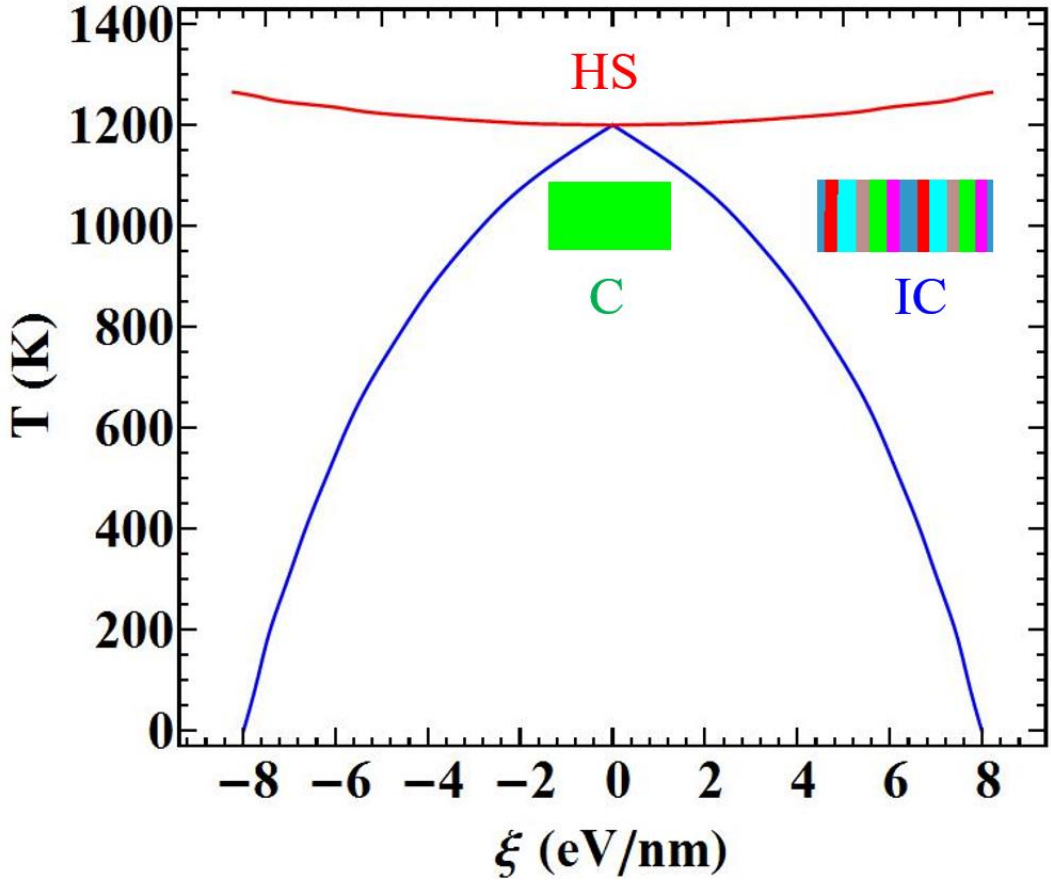

FIG. 5. Temperature-Lifshitz invariant coefficient (*T*-ξ) phase diagram of *h*-YMO. In the high-symmetry (HS) phase, all the order parameters equal to zero.

**F. Domain structure evolution from vortex domains to incommensurate phase**

In this section, we investigate how the domain structures evolve from vortex domains to the IC domains in YMO. The simulation grid is modified to $1024\Delta x \times 624\Delta x \times 1\Delta x$, which is a pseudo-2D system on the basal plane. To simulate the boundary condition with two free surfaces, an insulating layer with grid $1024\Delta x \times 400\Delta x \times 1\Delta x$ is added on top of the YMO layer, and the order parameters in the insulating layer are maintained as zero. Periodic boundary conditions are applied to the combined (YMO + insulating layer) system along three directions. The temperature is chosen at 1199 K, at which temperature the driving force from vortex domain to the IC phases is large.

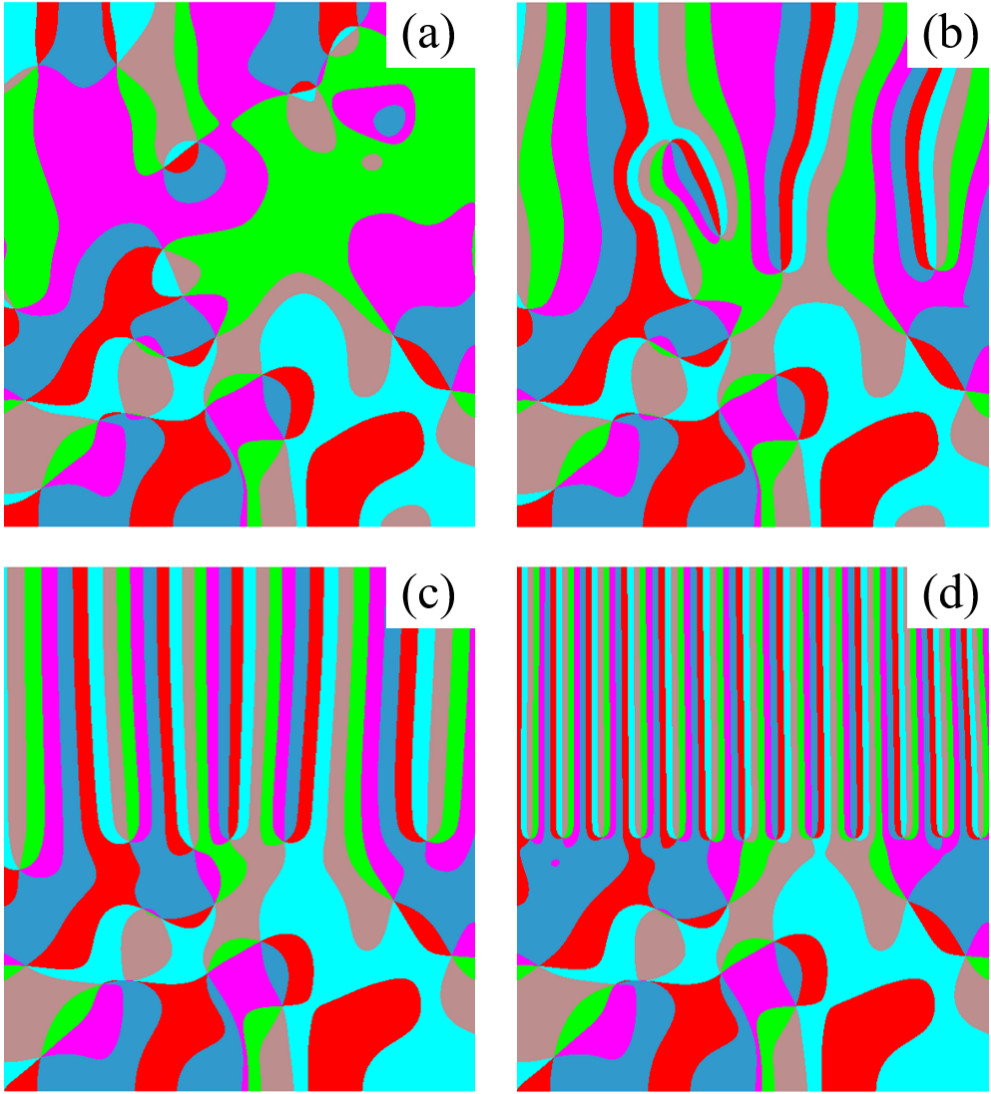

FIG. 6. Temporal evolution of the domain structures on the basal plane in *h*-YMO at 1199 K. (a) Initial domain structure. (b) Intermediate domain structure with $\xi$=1.49 eV/nm for the upper half, and $\xi$=0.00 eV/nm for the lower half. (c) Final domain structure with $\xi$=1.49 eV/nm for the upper half, and $\xi$=0.00 eV/nm for the lower half. (d) Final domain structure with $\xi$=2.98 eV/nm for the upper half, and $\xi$=0.00 eV/nm for the lower half. The colors are assigned based on the nearest C domains.

Fig. 6(a) shows a vortex domain structure, which is obtained starting from small random noises with $\xi$=0.00 eV/nm for the whole system [20]. Next we apply the external strain with $\xi$=1.49 eV/nm to the upper half while no strain is applied to the lower half. As shown in Fig. 6(b), in the upper half of the system, the vortices are pulled down whereas the anti-vortices are pulled up under the effect of the applied strain [10]. Finally, the anti-vortices are pulled out of the top surface of the system, and we obtain single-chirality striped domains in the upper half of the system (Fig. 6c). The vortices are lined up in the middle of the system, which form the IC-vortex boundary separating the IC domains and vortex domains. If the initial domain structure maintains the same as in Fig. 6(a), while the strain applied to the upper half is doubled to $\xi$=2.98 eV/nm, the final domain structure is given in Fig. 6(d), which shows a smaller wave length in the upper half than that in Fig. 6(c). Thus, a larger $\xi$ results in a smaller wave length of the IC domains, which is consistent with the conclusion of Fig. 4(b).

In Eq. (9), we only consider the interaction between strain and the gradient of the primary order parameter. The interaction with strain gradient, introduced in previous report to explain the vortex density at the IC-vortex boundary [10], is not included in this paper. In fact, Figs. 6(c) and 6(d) demonstrate that the vortex density at the IC-vortex boundary is determined by the value of $\xi$, i.e., by the magnitude of applied strain $(\varepsilon_{xx} - \varepsilon_{yy})$. This is because, as discussed below, the applied strain not only produces the Magnus-type force pulling the vortices and anti-vortices in opposite directions [10], but also leads to the creation and annihilation of vortex-antivortex pairs.

The detailed temporal evolution of domain structures with $\xi$=1.49 eV/nm for the upper half is demonstrated in Supplementary Movie I, and several zoomed-in snapshots are given in Figs. 7(a)-7(d). In Fig. 7, we label a vortex with an even number followed by a + sign, and an anti-vortex with an odd number followed by a – sign. As shown in Figs. 7(a) and 7(b), under the effect of the applied strain, a vortex-antivortex pair (3- and 4+) is created near two bubble-like domains. A typical process for the creation of a vortex-antivortex pair is sketched in Figs. 7(e)-7(g), i.e., the nucleation of two bubble-like domains within two neighboring domains, followed by the nucleation of another two domains. The phase-field simulation for the vortex-antivortex creation is shown in Movie II, which is the reverse process of the vortex-antivortex annihilation demonstrated in an earlier report [20]. The driving force for the domain nucleation is that in this case the self-energy of solitons is negative, which favors more solitons, i.e., more domain walls, as discussed in Section D. Also, the evolution from the vortex domains to the IC phase is accompanied by the annihilation of vortex-antivortex pairs, as shown in Figs. 7(b)-7(d). Through the creation and annihilation of vortex-antivortex pairs, the wave length of the IC phase in the final domain structure is close to the equilibrium wave length as indicated in Fig. 4(b), and independent of the initial vortex density. Therefore, the vortex density at the IC-vortex boundary is determined by the magnitude of the applied strain, and independent of the initial vortex density.

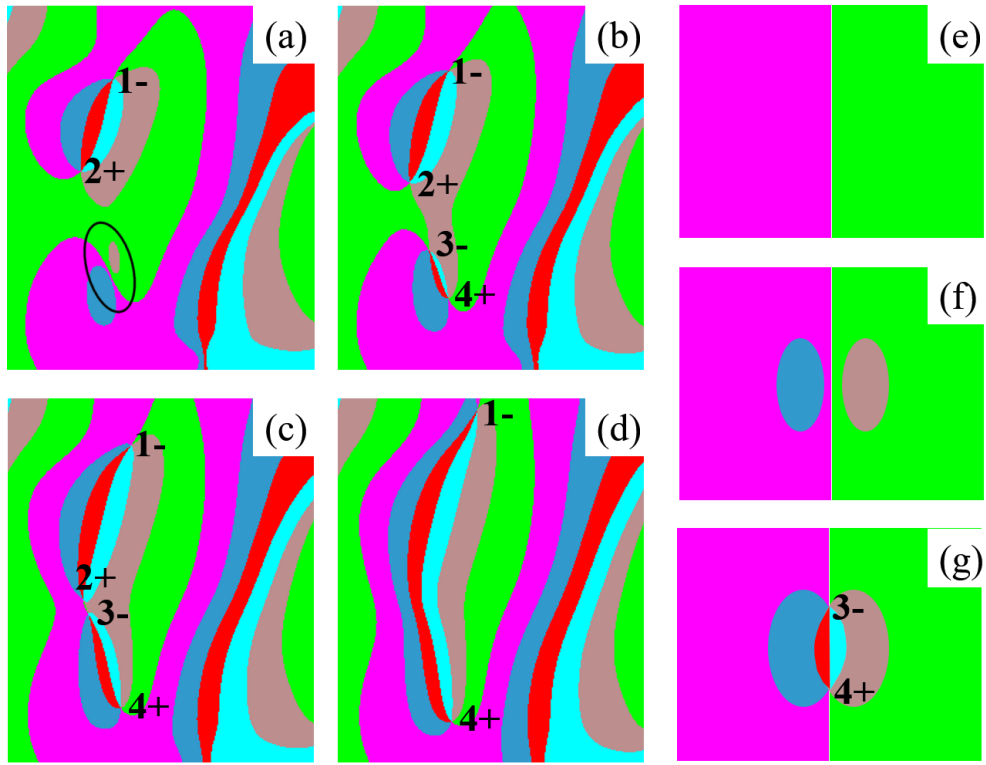

FIG. 7. Creation and annihilation of vortex-antivortex pairs under the effect of the applied strain. (a)-(d) Zoomed-in snapshots from a phase-field simulation. (a) and (b) show the creation of the vortex-antivortex pair (3- and 4+) near two bubble-like domains. The circled out region in (a) indicates the position where the vortex-antivortex pair is created. (b)-(d) demonstrate the annihilation of the vortex-antivortex pair (2+ and 3-). (e)-(g) Schematics for the creation of a vortex-antivortex pair from a domain wall.

## III. Conclusion

In summary, we investigate the stability and properties of the single-chirality stripe domains, i.e., the incommensurate phase, in hexagonal $REMnO_3$ (RE, rare earths) induced by applied strain. By comparing the numerical results of $REMnO_3$ with the analytical solution of the XY model, we demonstrate the effect of the energy anisotropy in the order parameter space. Surprisingly, in $REMnO_3$, when the applied strain is large, abnormal enhancement of the order parameter magnitude is observed near the transition regions between the plateaus corresponding to the commensurate domains. The equilibrium wave length is studied as a function of the applied strain at different temperatures, and a temperature-strain phase diagram is constructed for the stabilities of the incommensurate phase, commensurate phase, and high-symmetry phase. Phase-field simulations are employed to demonstrate the temporal evolution of the domain structures under the applied strain. It is found that the applied strain not only produces the force pulling the vortices and anti-vortices, but also results in the creation and annihilation of vortex-antivortex pairs. The study can serve as guidance for the manipulation and engineering of domains and associated topological defects in hard crystalline materials.

## Acknowledgements

The work is supported by the Penn State MRSEC, Center for Nanoscale Science, under the award NSF DMR-1420620 (FX) and by the U.S. Department of Energy, Office of Basic Energy Sciences, Division of Materials Sciences and Engineering under Award FG02-07ER46417 (FX and LQC). This work used the Extreme Science and Engineering Discovery Environment (XSEDE), which is supported by National Science Foundation grant number ACI-1548562 [31]. XW acknowledges the National Natural Science Foundation of China (Grant No. 11604011).

## References

[1]     C. Kittel, *Introduction to Solid State Physics* (Wiley, 2016), Eight edn.
[2]     J.-C. Tolédano and P. Tolédano, *The Landau theory of phase transitions: application to structural, incommensurate, magnetic and liquid crystal systems* (World Scientific Publishing Co Inc, 1987), Vol. 3.
[3]     P. Bak, Reports on Progress in Physics **45**, 587 (1982).
[4]     J. Axe, M. Iizumi, and G. Shirane, *Incommensurate phases in Dielectrics, ed. R. Blinc and AP Levanyuk* (North-Holland, Amsterdam, 1986).
[5]     H. Z. Cummins, Physics Reports **185**, 211 (1990).
[6]     J. A. Wilson, F. Di Salvo, and S. Mahajan, Advances in Physics **24**, 117 (1975).
[7]     R. Coleman, B. Giambattista, P. Hansma, A. Johnson, W. McNairy, and C. Slough, Advances in Physics **37**, 559 (1988).
[8]     A. Morozovska, E. Eliseev, J. Wang, G. Svechnikov, Y. M. Vysochanskii, V. Gopalan, and L.-Q. Chen, Physical Review B **81**, 195437 (2010).
[9]     L. Jiang, Y. Zhou, Y. Zhang, Q. Yang, Y. Gu, and L.-Q. Chen, Acta Materialia **90**, 344 (2015).
[10]    X. Wang, M. Mostovoy, M.-G. Han, Y. Horibe, T. Aoki, Y. Zhu, and S.-W. Cheong, Physical review letters **112**, 247601 (2014).
[11]    T. Choi, Y. Horibe, H. Yi, Y. Choi, W. Wu, and S.-W. Cheong, Nature materials **9**, 253 (2010).
[12]    S. Chae, Y. Horibe, D. Jeong, S. Rodan, N. Lee, and S.-W. Cheong, Proceedings of the National Academy of Sciences **107**, 21366 (2010).
[13]    C. J. Fennie and K. M. Rabe, Physical Review B **72**, 100103 (2005).
[14]    B. B. Van Aken, T. T. Palstra, A. Filippetti, and N. A. Spaldin, Nature materials **3**, 164 (2004).
[15]    A. S. Gibbs, K. S. Knight, and P. Lightfoot, Physical Review B **83**, 094111 (2011).

[16] S. Artyukhin, K. T. Delaney, N. A. Spaldin, and M. Mostovoy, Nature materials **13**, 42 (2013).
[17] P. M. Chaikin and T. C. Lubensky, *Principles of condensed matter physics* (Cambridge Univ Press, 2000), Vol. 1.
[18] L. Chen and J. Shen, Computer Physics Communications **108**, 147 (1998).
[19] L.-Q. Chen, Annual review of materials research **32**, 113 (2002).
[20] F. Xue, X. Wang, Y. Gu, L.-Q. Chen, and S.-W. Cheong, Scientific reports **5**, 17057 (2015).
[21] A. Levanyuk, S. Minyukov, and A. Cano, Physical Review B **66**, 014111 (2002).
[22] A. Tagantsev, Ferroelectrics **69**, 321 (1986).
[23] G. Rupprecht and R. Bell, Physical Review **135**, A748 (1964).
[24] S.-Z. Lin *et al.*, Nature Physics **10**, 970 (2014).
[25] S. M. Griffin, M. Lilienblum, K. Delaney, Y. Kumagai, M. Fiebig, and N. Spaldin, Physical Review X **2**, 041022 (2012).
[26] J. Li, F.-K. Chiang, Z. Chen, C. Ma, M.-W. Chu, C.-H. Chen, H. Tian, H. Yang, and J. Li, Scientific Reports **6** (2016).
[27] A. Jacobs and M. Walker, Physical Review B **21**, 4132 (1980).
[28] A. Y. Borisevich *et al.*, Nature communications **3**, 775 (2012).
[29] S. Chae, N. Lee, Y. Horibe, M. Tanimura, S. Mori, B. Gao, S. Carr, and S.-W. Cheong, Physical review letters **108**, 167603 (2012).
[30] C. M. Lapilli, P. Pfeifer, and C. Wexler, Physical review letters **96**, 140603 (2006).
[31] J. Towns *et al.*, Computing in Science & Engineering **16**, 62 (2014).